\def\kms{$\rm km\;s^{-1}$}

\documentstyle[11pt,aaspp4]{article} 
\slugcomment{{\it to appear in The Astrophysical Journal Letters}}

\lefthead{Bertola et al.}
\righthead{The bulge-disk orthogonal decoupling in galaxies: NGC~4698}

\begin{document}

\title{The Bulge-Disk Orthogonal Decoupling in Galaxies: NGC~4698\footnotemark}

\footnotetext{Based on observations carried out at ESO, La Silla (Chile)
(ESO N. 60.A-0800)}

\author{F.~Bertola\altaffilmark{2},
E.M.~Corsini\altaffilmark{2},
J.C.~Vega Beltr\'an\altaffilmark{3},
A.~Pizzella\altaffilmark{4},
M.~Sarzi\altaffilmark{2},
M.~Cappellari\altaffilmark{2},
and J.G.~Funes,~S.J.\altaffilmark{2}}

\altaffiltext{2}{Dipartimento di Astronomia, Universit\`a di Padova,
Vicolo dell'Osservatorio 5, I-35122 Padova, Italy}
\altaffiltext{3}{Telescopio Nazionale Galileo, Osservatorio Astronomico
di Padova, Vicolo dell'Os\-servatorio 5, I-35122 Padova, Italy}
\altaffiltext{4}{European Southern Observatory, Alonso de Cordova 3107,
Casilla 19001, Santiago 10, Chile}

\begin{abstract}

The $R-$band isophotal map of the Sa galaxy NGC~4698 shows that the
inner region of the bulge structure is elongated perpendicularly to
the major axis of the disk, this is also true for the outer parts of
the bulge if a parametric photometric decomposition is adopted. At
the same time the stellar component is characterized by an inner
velocity gradient and a central zero-velocity plateau along the minor 
and major axis of the disk respectively. This remarkable geometric
and kinematic decoupling suggests that a second event occurred in the
formation history of this galaxy.
 
\end{abstract}

\keywords{galaxies: individual (NGC~4698) --- galaxies: kinematics and
dynamics --- galaxies: spiral --- galaxies: structure --- galaxies:
formation}

\section{Introduction}

In the course of an investigation of the kinematic properties of
early-type spiral galaxies we encountered the peculiar case of
NGC~4698, which is the subject of this paper.

This galaxy is characterized by a remarkable geometric decoupling
between bulge and disk, whose apparent major axes appear oriented in
an orthogonal way at a simple visual inspection of the galaxy images
(e.g., see Panels 78, 79 and 87 in Sandage \& Bedke 1994).  The
stellar rotation curve exhibits an unusual zero-velocity plateau in
the central portion of the disk major axis. A spectrum taken along the
disk minor axis shows an inner velocity gradient of the stellar
component suggesting the presence of a kinematic decoupling between
the inner regions of the galaxy and its disk. This decoupling is a
direct indication that distinct events occurred in the history of
NGC~4698.  For this reason it should be considered a noteworthy case
for the interpretation of the processes leading to the formation of
spirals.

NGC~4698 is classified Sa by Sandage \& Tammann (1981) and Sab(s) by
de Vaucouleurs et al. (1991, RC3).  Sandage \& Bedke (1994, CAG) in The
Carnegie Atlas of Galaxies presented NGC~4698 in the Sa section
as an example of the early-to-intermediate Sa type.  They describe the
galaxy as characterized by a large central E-like bulge in which there
is no evidence of recent star formation or spiral structure. The
spiral arms are tightly wound and become prominent only in the outer
parts of the disk. They are defined primarily by the dust which forms
fragmentary lanes of the multiple-armed type. NGC~4698 possesses all
the characteristics of a spiral galaxy and therefore it is
morphologically different from the ellipticals with polar ring,
although in both cases we are facing a similar phenomenon of
orthogonal geometric and kinematic decoupling (e.g. AM~2020-504 in
Whitmore et al. 1990).

NGC~4698 belongs to the Virgo cluster.  Its total $B-$band
magnitude is $B_T=11.46$ mag (RC3) which corresponds to $M_B=-19.69$
mag assuming a distance of 17 Mpc (Freedman et al. 1994).

\section{Observations and data reduction}

The photometric and spectroscopic observations of NGC~4698 were
carried out with EFOSC2 at the ESO 3.6-m telescope in La Silla on
March 1998, 20-21.

We obtained three 20-seconds images of the galaxy using the No.~40
Loral/Lesser $2048\,\times\,2048$ CCD with $15\,\times\,15$ $\mu$m$^2$
pixels in combination with the No.~642 $R-$band filter. It yielded an
unvignetted field of view of $3\farcm8\,\times\,5\farcm3$ with an
image scale of $0\farcs32$ pixel$^{-1}$ after an on-line binning of
$2\times2$ pixels.  The routine data reduction has been carried out
using IRAF. Gaussian fit to the field stars in the final processed
image yielded a point spread function FWHM$\,=\,1\farcs0$.  The
absolute calibration was made using the multi-aperture photometric
data obtained in the same band by Schr\"oeder \& Visvanathan (1996)
since no photometric standards were observed.

Two 30-minutes spectra were taken along the major ($\rm
P.A.\,=\,170^\circ$) and the minor axis ($\rm P.A.\,=\,80^\circ$) of
the galaxy. We used the No.~8 grism with 600 $\rm grooves\;mm^{-1}$ in
combination with a $1\farcs0\,\times\,5\farcm7$ slit. The No.~40 CCD
was adopted as detector.  They yielded a wavelength coverage of 2048
\AA\ between 4320 \AA\ and 6368 \AA\ with a reciprocal dispersion of
66.33 $\rm \AA\,mm^{-1}$. Each pixel of the spectra corresponds to
$\rm 1.99\,\AA\,\times\,0\farcs32$ after an on-line binning of
$2\,\times\,2$ pixels. Comparison lamp exposures were obtained before
and after each object integration. Some spectra of late-G or early-K
giant stars were taken to serve as template in measuring the stellar
kinematics. The seeing FWHM during the observing nights was between
$1''$ and $1\farcs5$ as measured by the Differential Image Motion
Monitor of La Silla. Using standard MIDAS routines the spectra were
bias subtracted, flat-field corrected, and wavelength
calibrated. Cosmic rays were found and corrected by comparing the
counts in each pixel with the local mean and standard deviation and
substituting a suitable value. The instrumental resolution was derived
by measuring the FWHM of a dozen of single emission lines distributed
all over the spectral range of a calibrated comparison spectrum. It
corresponds to a $\rm FWHM = 6.41 \pm\ 0.07$
\AA\ (i.e. $\sigma = 124$ \kms\ at H$\alpha$).  The single spectra
obtained along the same axis were aligned and co-added using their
stellar-continuum centers as reference. In each spectrum the center of
the galaxy was defined as the center of the Gaussian fitting the
radial profile of the stellar continuum.  The contribution of the sky
was determined from the edges of the resulting frame and then
subtracted. The stellar kinematics was measured from the spectra
absorption lines using the Fourier Correlation Quotient Method (Bender
1990) as done by Bertola et al. (1996).

\section{Results}

\subsection{Photometry}

The $R-$band isophotal map of NGC~4698 is presented in Fig.~1. A
geometric decoupling ($\rm \Delta\,P.A.\,\simeq\,90^\circ$) is visible
both in the inner isophotes (see inset) and in the outermost one,
which is characterized by two `bumps' oriented perpendicularly to the
galaxy major axis.  The isophotes between $4''$ and $18''$ appear
round in the plot. However as soon as an exponential disk is
subtracted they also become elongated perpendicularly to the disk
major axis (Fig.~3).  The overall shape of the isophotes between
$18''$ and $32''$ is similar to the one observed in $V$ (Takase,
Kodaira \& Okamura 1984) in $B$ (Yasuda, Okamura \& Fukugita 1995) in
$K'$ (Boselli et al. 1997) and in $K-$band (Moriondo, Giovanardi \&
Hunt 1998) images. This suggests that the dust does not play any role in
shaping the galaxy isophotes.

We performed a bidimensional photometric decomposition of the NGC~4698
surface-brightness, which we assumed to be the sum of a $r^{1/4}$
bulge and an exponential disk.  To avoid the effects of the dust lanes
we performed the decomposition on the image obtained by folding the
eastern side of NGC~4698 around the galaxy major axis. In performing
the decomposition the seeing convolution has been taken into account.
The best-fit parameters are $\mu_e = 20.4$ mag$\cdot$arcsec$^{-2}$,
$r_e = 21\farcs0$ and $q\,=\,(b/a)_{\it bulge}\,=\,1.14$ (where $a$
and $b$ are taken along the direction of the galaxy major and minor
axis respectively) for the bulge and $\mu_0 = 19.6$
mag$\cdot$arcsec$^{-2}$, $r_d = 42\farcs9$, and $i={\rm
arccos}\,(b/a)_{\it disk}\,=\,65^\circ$ for the disk.  The fact that
the axial ratio of the bulge is found to be greater than unity
confirms the exceptional property of NGC~4698 of having a bulge
elongated along the disk minor axis. The surface-brightness profiles
derived along the major and minor axes of NGC~4698 and the
corresponding bulge-disk decomposition are presented in Fig.~2.

The NGC~4698 residual image obtained by subtracting the surface
brightness of the exponential disk derived in the photometric
decomposition is shown in Fig.~3.

The difference between the parametric modeled and the measured
surface-brightness profiles which results between $14''$ and $32''$
along the galaxy major axis (Fig.~2) could be produced by a local
fluctuation of the light distribution of the disk.

\subsection{Kinematics}

The stellar velocity curve measured along the major axis of NGC~4698
is characterized by a central plateau, indeed the stars have a zero
rotation for $|r|\,\leq\,8''$ ($0.7$ kpc).  At larger radii the
observed stellar rotation increases from zero to an approximately
constant value of about 200 \kms\ for $|r|\,\gtrsim\,50''$ ($4.1$ kpc)
up to the farthest observed radius at about $80''$ ($6.6$ kpc).  The
stellar velocity dispersion profile has been measured out to $30''$
($2.5$ kpc). It is peaked in the center at the value of 185 \kms . The
stellar velocities measured by Corsini et al. (1999) agree within the
errors with the data of this paper. The ionized gas kinematics along
the major axis of NGC~4698 has been studied by Rubin et al. (1985) and
by Corsini et al. (1999).  We measured the minor-axis stellar
kinematics out to about $20''$ on both sides of the galaxy.  In the
nucleus the stellar velocity rotation increases to about 30 \kms\ at
$|r|\,\simeq\,2''$. Then it decreases between $2''$ and $6''$ and it
is characterized by an almost zero value beyond $6''$.  The velocity
dispersion profile has a central maximum of 175
\kms\ in agreement within the errors with the value measured along the
major axis. The velocity curves and the velocity-dispersion radial
profiles of the stellar component (out only to $28''$ for the spectrum
along the major axis) are shown in Fig.~4.

The spectral resolution and signal-to-noise ratio of our spectra were
not sufficient to perform a two-component kinematic decomposition via
double-Gaussian fit of the line-of-sight velocity distribution (LOSVD)
to disentangle the contribution of bulge and disk to the observed
major and minor-axis kinematics.  In order to reproduce the observed
velocity curve along the galaxy major axis we modeled the observed
LOSVD in the following way. For the bulge we assumed a constant zero
velocity and a constant velocity dispersion of $\sigma_{\it
bulge}\,=\,180$ \kms\ at all radii, while for the disk we took a velocity 
rising linearly to match the outer points of the plotted curve (where the
light contribution of the bulge is negligible) with a constant
velocity dispersion of $\sigma_{\it disk}\,=\,100$ \kms.  The
resulting velocity curve obtained by fitting with a Gaussian the sum
of the two Gaussian components of bulge and disk weighted according to
the photometric decomposition of Fig.~2 is shown with a continuous
line in the upper panel of Fig.~4.  The agreement with the observed
points is good. In particular the flat central part of the observed
velocity curve is well reproduced. For $|r|>50''$ the bulge
contribution to the galaxy light is negligible and the constant
stellar rotation ($V\,\simeq\,200$ \kms) we measure out to $80''$ can
be directly explained by the differential rotation of the disk.

As far as the kinematics along the minor axis is concerned, the
observed radial velocities and velocity dispersions are to be ascribed
mainly to the bulge component, which dominates over the disk at all
radii.  It results that the rotation of the bulge is mainly limited to
the central region.  For $|r|>6''$ the lack of detailed information
about the shape of the LOSVD prevents us to detect even a small
overall rotation of the bulge, in spite of what one would expect
considering the elongated shape of its isophotes. Therefore it is
reasonable to ascribe to the entire bulge a projected angular momentum
perpendicular to that of the disk.

\section{Discussion and conclusion}

\subsection{The bulge-disk decoupling scenario}

In the previous paragraph adopting a parametric photometric
decomposition, we have pointed out that the bulge and disk of the Sa
NGC~4698 appear on the sky elongated perpendicularly to each other and
characterized by the rotation around two orthogonal axes.  Assuming
that the intrinsic shape of bulges is generally triaxial (Bertola,
Vietri \& Zeilinger 1991) and that the plane of the disk coincides
with the plane perpendicular either to the bulge major or minor axis,
we deduce from the observed configuration that the major axis of the
bulge is perpendicular to the disk, given that the latter is seen not
far from edge on. The fact that the velocity field of the bulge is
characterized by a zero velocity along its apparent minor axis (as
indicated by the central plateau in the rotation curve along the disk
major axis) and by a velocity gradient along its major axis suggests
that the rotation axis of the bulge lies on the plane of the disk, and
therefore the intrinsic angular momenta of bulge and disk are
perpendicular.

The orthogonal decoupling of the bulge and disk in NGC~4698 indicates
that a second event occurred in the formation history of this
galaxy. We suggest that the disk has formed at a later stage due to
the acquisition of material by a triaxial spheroid on its principal
plane perpendicular to the major axis. An example of acquisition on
the plane perpendicular to the minor axis could be represented by
NGC~7331 (Prada et al. 1996), where the bulge has been found
counterrotating with respect to the disk (although this result has
been recently questioned by Bottema 1999). Up to now NGC~4698 and
NGC~7331 represent the only cases of kinematic evidence that disk
galaxies with prominent bulges could be started as `undressed
spheroids' and their disks accreted gradually over several billion
years, as suggested by Binney \& May (1986). Recently such kind of
processes have been considered within semi-analytical modeling
techniques for galaxy formation, where the disks accrete around bare
spheroids previously formed either directly from the relaxation of gas
in a spherical distribution parallel to that of their surrounding dark
halos (Kauffmann 1996), or from the merging of disk proto-galaxies
previously formed (Baugh, Cole \& Frenk 1996).  In this framework we
could expect that the shape of the dark matter halo of NGC~4698
correlates with that of the pre-existing galaxy (namely the actual
bulge) as in the case of the prototype polar-ring galaxy NGC~4650A. In
this galaxy Sackett et al. (1994) found a highly flattened dark halo,
which has the same orientation of the pre-existing S0 rather than that
of the acquired gaseous ring. If the disk of NGC~4698 has an external
origin then polar-ring elliptical galaxies like NGC~5266 (Varnas et
al. 1987) and AM~2020-504 (Whitmore et al. 1990) and ellipticals with
dust lane along the minor axis (Bertola 1987) could represent
transient stages towards the formation of spiral systems like
NGC~4698.

The possible presence of an end-on bar lying in the plane of the disk
is ruled out since we measure along the major axis a zero-velocity
plateau instead of a strong velocity gradient similar to those
predicted by Merrifield (1996) for end-on bars in edge-on galaxies.

\subsection{An alternative scenario}

A non-parametric photometric decomposition of the NGC~4698
surface-brightness distribution has been recently proposed by Moriondo
et al. (1998). The resulting bulge shows circular isophotes and it
dominates the galaxy light out to about $20''$ from the center along
the galaxy major axis, the surface-brightness profile of the disk
flattens in the inner $30''$.

This decomposition removes the oval features at the two sides of the
bulge in the disk-subtracted image of NGC~4698 (Fig.~3) which
correspond to the light excess observed between $14''$ and $32''$
along the major axis of the parametric modeled profile (Fig.~2). The
disk light contribution becomes so small for $|r|<15''$ along the
major axis that the rotation we observe has to be attributed mainly to
the bulge component, which indeed would rotate in the same sense of
the disk. In order to explain the central plateau along the disk major
axis, the velocity gradient and the elongation of the inner isophotes
along the disk minor axis it is necessary to assume the presence of a
third luminous component in the central region of NGC~4698 in addition
to the round bulge and the disk with a flattened surface-brightness
profile. It should be noted that in the statistics of Bertola et
al. (1991) round bulges (as well as those perpendicular to the disk)
are not present.

If we adopt the non-parametric photometric decomposition we are lead
to conceive the presence in the center of NGC~4698 of a structure
similar to that of the kinematically decoupled cores observed in
several ellipticals (see Mehlert et al. 1998 for a list). In
ellipticals isolated cores generally tend to not show up
photometrically (de Zeeuw \& Franx 1991) while the isophotes of the
NGC~4698 core is characterized by an orthogonal geometric decoupling
with respect to those of the disk. Therefore the core in NGC~4698
(which should be the first case of an isolated core observed in a
spiral galaxy) is photometrically decoupled from the disk and
kinematically decoupled with respect to both bulge and disk. Also
according to this alternative scenario NGC~4698 experienced a second
event in its history.

\acknowledgments
We thank H.-W. Rix for useful discussions.
JCVB acknowledges a grant from Telescopio Nazionale Galileo and Osservatorio
Astronomico di Padova.

\clearpage
\plotone{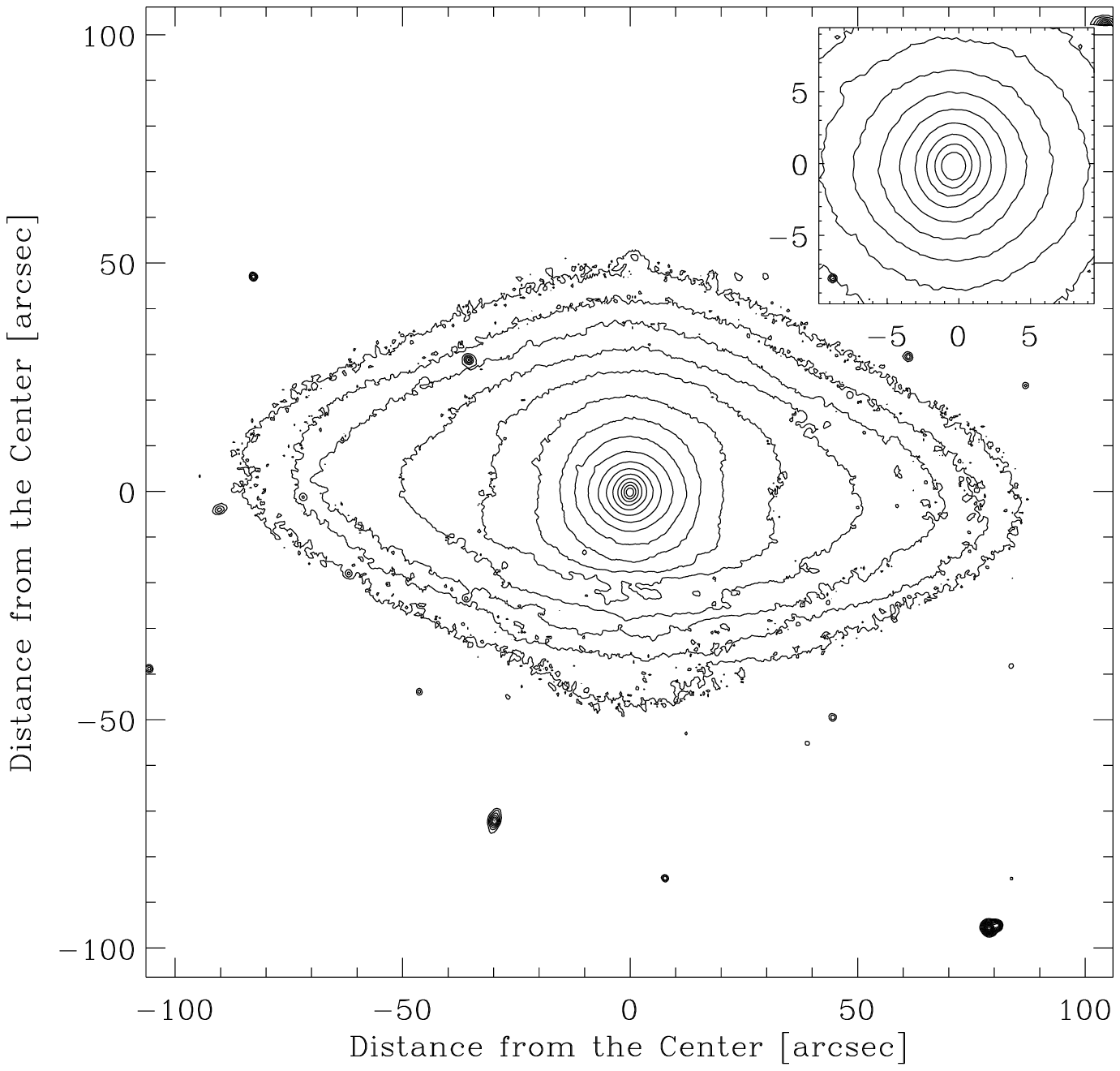}
\figcaption{The $R-$band isophotes of NGC~4698 (boxcar smoothed over
3$\times$3 pixels). Isophotes are given in steps of 0.4
mag$\cdot$arcsec$^{-2}$ with the outermost one corresponding to 21.8
mag$\cdot$arcsec$^{-2}$ and the central one to 15.8
mag$\cdot$arcsec$^{-2}$. In the {\it inset\/} the (non smoothed) isophotal
map of the inner $10''$ is plotted. North is right and east up}

\clearpage
\plotone{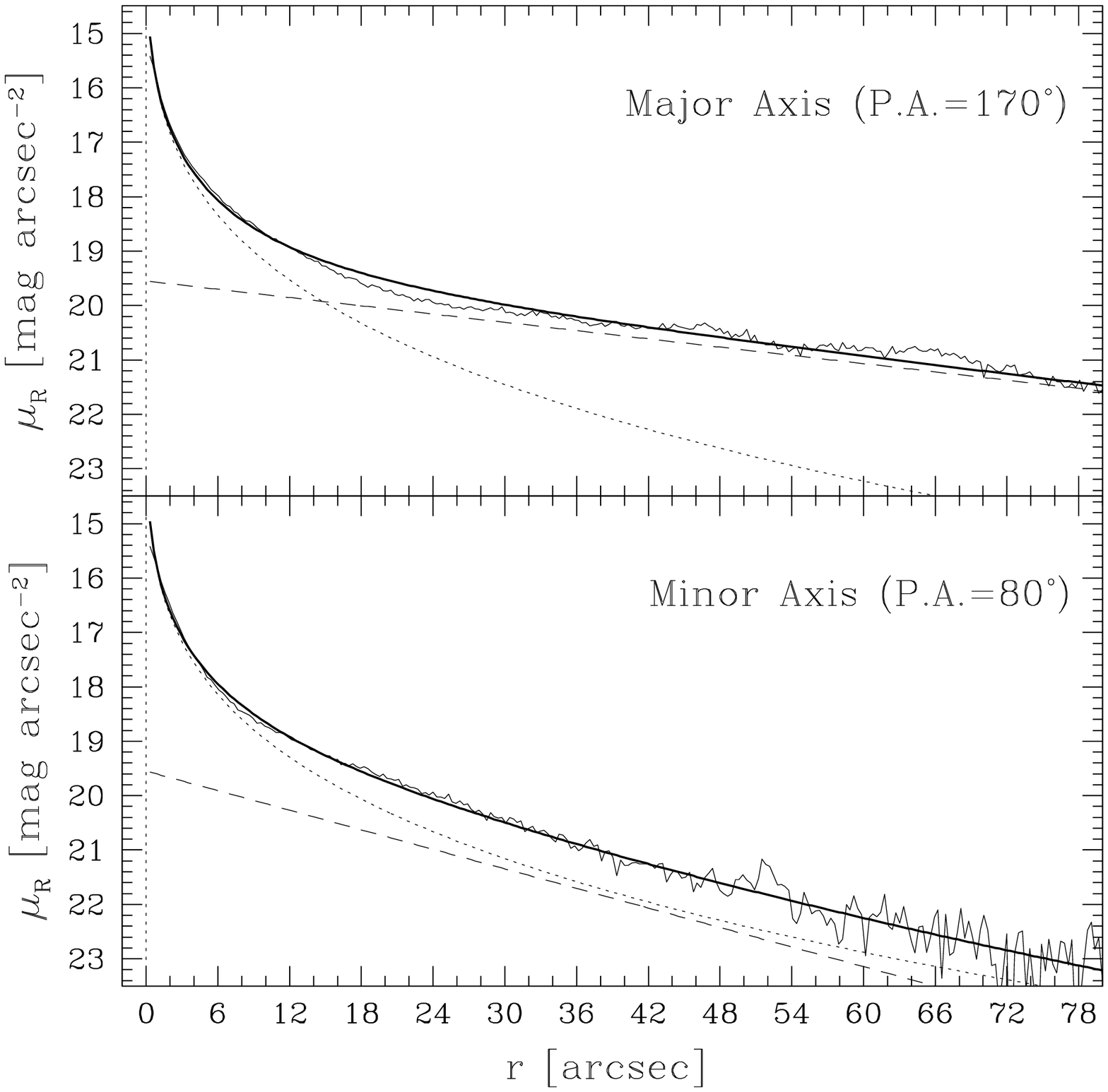}
\figcaption{The $R-$band surface-brightness profiles of NGC~4698 ({\it
thin continuous line\/}) along the major ($\rm
P.A.\,=\,170^\circ$) and minor axis ($\rm P.A.\,=\,80^\circ$) of
the galaxy out to $80''$ from the center.  The surface-brightness
profiles of the $r^{1/4}$ bulge ({\it dotted line\/}), the
exponential disk ({\it dashed line\/}) and their sum 
({\it thick continuous line\/}) are also plotted}

\clearpage
\plotone{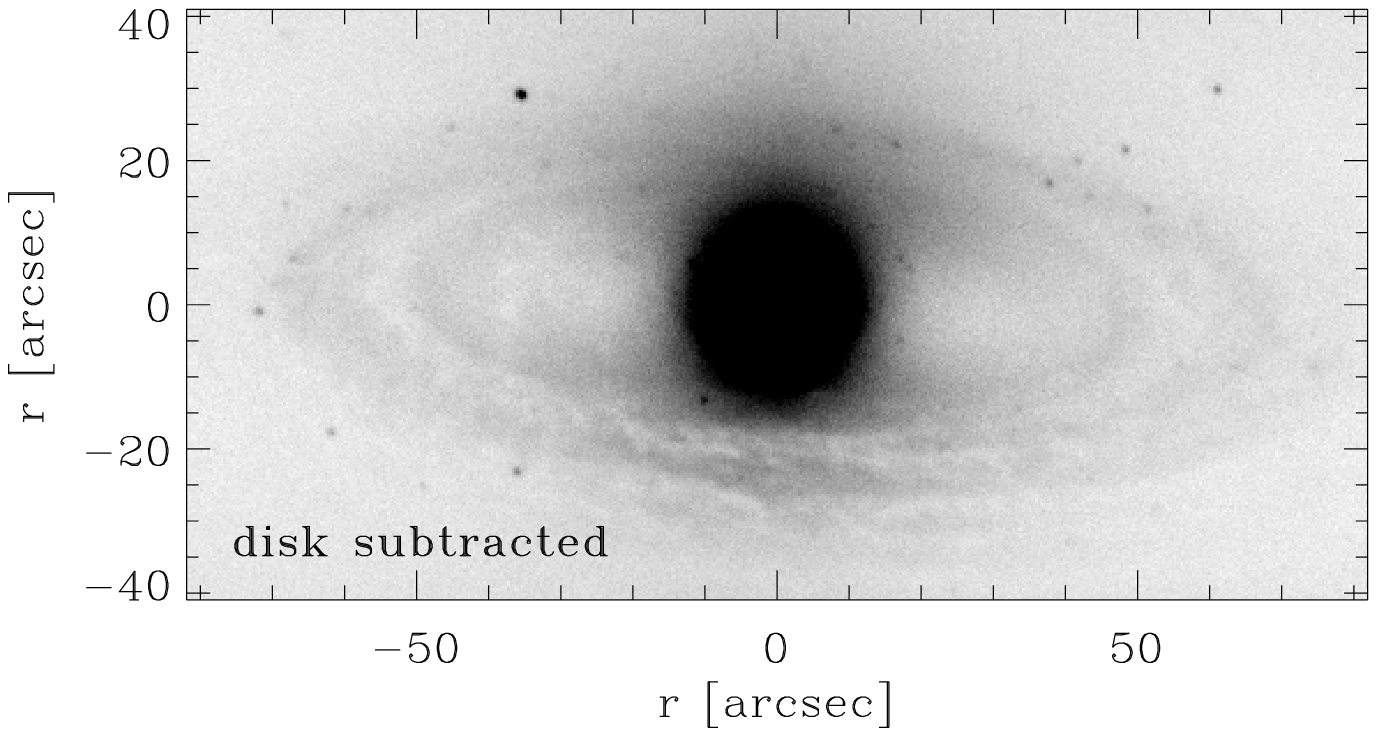}
\figcaption{The $R-$band residual image of NGC~4698 obtained after the
subtraction from the total surface-brightness distribution of the
light contribution due to the exponential disk of the photometric
bidimensional decomposition. North is right and east up}

\clearpage
\plotone{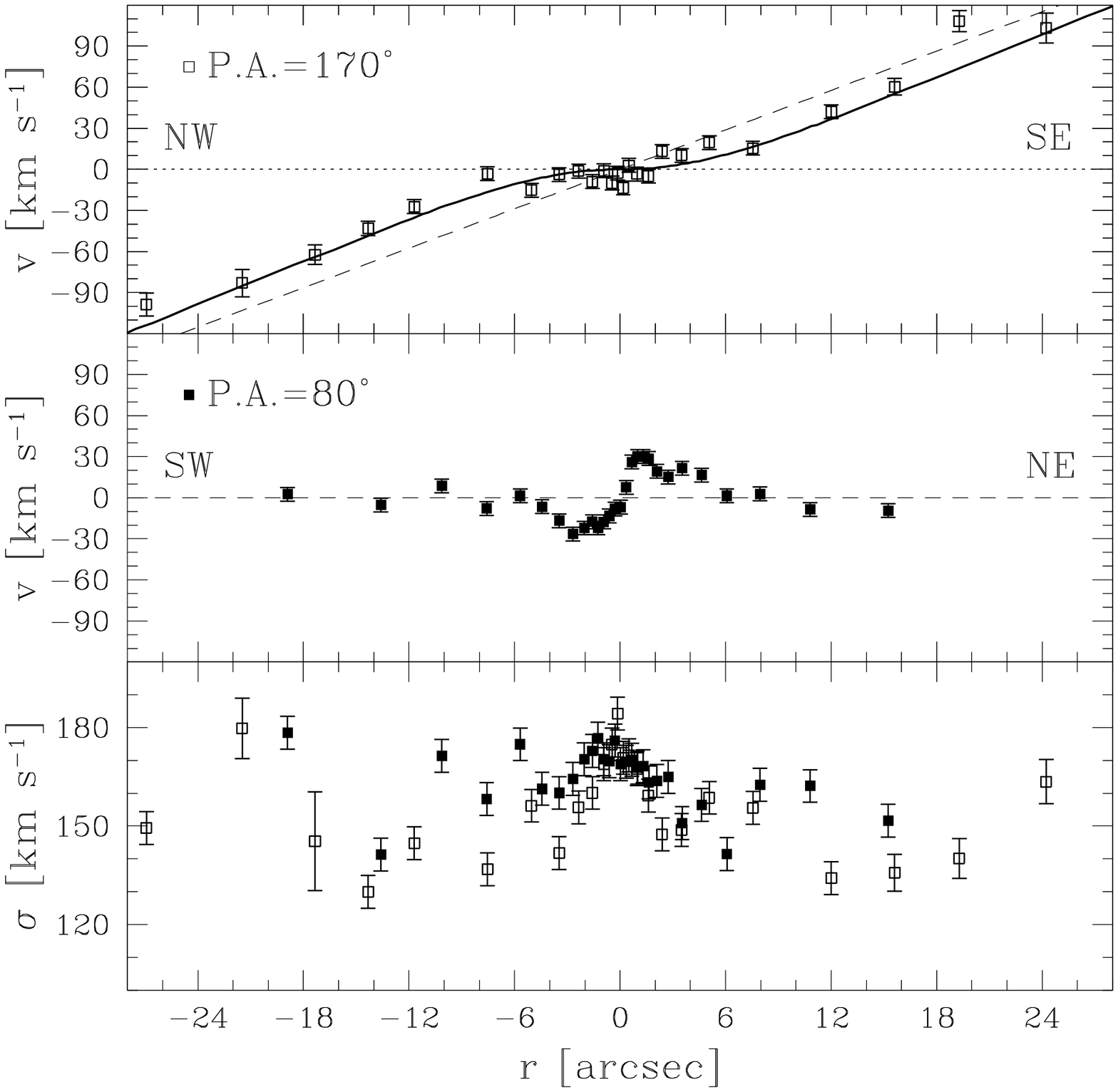}
\figcaption{The observed stellar rotation velocity and velocity dispersion
as a function of radius along the major ({\it open squares\/}) and
minor axis ({\it filled squares\/}) of NGC~4698 out to $28''$ from the
center. The heliocentric system velocity is $V_\odot = 992 \pm 10$
\kms. The {\it dotted\/} and {\it dashed lines\/} represent the
velocity contribution of the bulge and disk components to the total
velocity ({\it thick continuous line\/}) of our model}

\end{document}